\documentclass[11pt]{article}
\usepackage[dvips]{graphicx}
\usepackage{a4wide}
\usepackage{amsmath}
\usepackage{amsthm}
\usepackage{amsfonts}

{\theoremstyle{plain}%
  \newtheorem{thm}{\bf Theorem}[section]%
  \newtheorem{lem}[thm]{\bf Lemma}%
  \newtheorem{prop}[thm]{\bf Proposition}%
  \newtheorem{cor}[thm]{\bf Corollary}%
}
{\theoremstyle{definition}
  
}
{\theoremstyle{remark}
  
}

\newcommand{\dC}{{\mathbb{C}}}%
\newcommand{\dN}{{\mathbb{N}}}%
\newcommand{\dQ}{{\mathbb{Q}}}%

\title{A Characterization of the Graphs to Induce Periodic Grover Walk}
\author{Yusuke Yoshie\footnote{Graduate School of Information Sciences, Tohoku University, Aoba, Sendai 980-8579, Japan (E-mail: yusuke.yoshie.r1@dc.tohoku.ac.jp) }\\}
\date{}

\begin {document}
\maketitle{}
\begin{abstract}
Recently, a research on quantum walks has been developed in various areas. In this paper we focus on the periodicity of the Grover walk which is one of quantum walk on the discrete graphs. Then we found some special graphs to induce a periodic Grover walk: At some time $k$, the quantum state $\varphi _{k}$ returns its initial quantum state $\varphi _{0}$. Our purpose is to characterize graphs which induce a $k$-periodic Grover walk for a fixed integer $k$. We do it for $k=2, 3, 4, 5$ and obtain a necessary condition for odd $k$.
\end{abstract}
\section {Introduction}

\subsection {Background and Notation}

Quantum walks (QWs) were introduced as a quantization of random walks (RWs) \cite{Reit11}. QWs are determined by a graph, its induced Hilbert space $\mathcal{H}$ and a unitary time evolution operator on $\mathcal{H}$. The amplitude of quantum walker is obtained by this unitary iteration to some initial state. Due to the unitarity of the time iteration, the norm of the amplitude is preserved, which implies that the distribution can be defined at each time step. However it is believed that there are no non-trivial representations of the present distribution of QWs by that of past time like Markov chain \cite{gud}. QWs are applied to various study fields, for example, a problem of searching marked elements on graphs \cite{Sze04}, \cite{Amba07}, \cite{Mag07}, \cite{Mag05}, fundamental physics \cite{Stra07}, \cite{Chisa11}, the limit theorems for its statistical behaviors \cite{Konno08}, \cite{Luczak}, spectral analysis \cite{Cante10}, \cite{Konno11}, and photon synthesis \cite{Lloy08}. In \cite{Sze04}, Szegedy walk was formulated as a natural quantization of reversible Marcov chains, and Szegedy showed that in most cases the quantized walk hits the marked set within the  square root of the classical hitting time. Grover walk, which is a special case of Szegedy walk is widely studied quantum walk. It is related to the analysis of zeta function and isomorphic problem between two cospectrum strongly regular graphs \cite{HSegawa13}, \cite{Sato11}. In \cite{Gro96}, Grover's search algorithm was introduced to search marked elements in database. 

In this paper we focus on the periodicity of the Grover walk on graphs. The periodicity means that the quantum state at some time returns to the initial state. Recently, the periodicity of QWs has been studied. In \cite{Konno13}, Konno, Takei and Shimizu studied Hadamard walk on cycle graph $C_{n}$. They found that only $C_{2}, C_{4}$, and $C_{8}$ induce periodic Hadamard walks, whose periodicity is $2, 8$, and $24$, respectively. In \cite{Higuchi13}, Higuchi, Konno, Sato and Segawa studied some QWs on several finite graphs. They found the condition of these graphs to induce periodic QWs. Their results are as follows:
\begin{itemize}
\item (Complete graphs) Szegedy walks induced by isotropic random walk with laziness $l$ with $l \in [0, 1] \cap \dQ$ on $K_{n}$ are periodic if and only if $(n,l)=(2,0), (3,0), (n,1/n), (2,1/4)$, or $(n,(n+1)/(2n))$, whose period is $2, 3, 4, 6$ and $6$, respectively.
\item (Complete bipartite graphs) Szegedy walks induced by isotropic random walk with laziness $l$ with $l \in [0, 1] \cap \dQ$ on $K_{n,m}$ with $m, n > 0$ and $m+n\geq 3$ are periodic if and only if $l=0,$ or $1/2$, whose period is $4$ and $12$, respectively.
\item (Strongly regular graphs) Grover walks on strongly regular graphs with parameters $(n, k, \lambda, \mu )$ are periodic if and only if \[ (n, k, \lambda , \mu )=(2k,k,0,k), (3\lambda ,2\lambda ,\lambda ,2\lambda), (5,2,0,1), \]
whose period is $4, 12$ and $5$, respectively.
\item (Cycle graphs) Szegedy walks induced by random walk with non-revertible probability $p$ with which walker moves clockwise on $C_{n}$ are periodic if and only if $p=(2-\sqrt{3})/4, (2-\sqrt{2})4$, $1/4$ for $n=2$, whose period is $6, 8$ and $12$, respectively, or $p=(2-\sqrt{3})/4, (2-\sqrt{2})4$, $1/4$ for $n=4$, whose period is $12, 8$ and $12$, respectively, or $p=(2-\sqrt{2})/4$ for $n=8$, whose period is $24$. 
 \end{itemize}
On graphs which induce a periodicity, the distribution iterates with some period, which implies that there are no limit distributions of QW. So we can say that such graphs are special class of graphs from the viewpoint of QWs. Our purpose is to characterize such special class of graphs. In the previous results, for fixed graphs, they found the condition of graphs to induce periodic QWs. On the contrary, we fixed an integer $k$ and tried to characterize graphs to induce a $k$-periodic Grover walk. 

We give $\dC^{|D(G)|}$ as $\mathcal{H}$, and the $|D(G)| \times |D(G)|$ unitary operator called the Grover transfer matrix as $U$, where $D(G)$ is a set of symmetric arcs of a finite graph $G$. A walker of Grover walk on graphs transfers on arcs. A motion of the quantum walker is interpreted as a dynamics of plane wave on the metric graphs \cite{qua}.

First, we introduce the notations and QWs on discrete graphs. All graphs $G$ considered in this paper are finite and simple graph without loops and multiple edges. Let $V(G)$, $E(G)$ be a set of vertices and edges of $G$ and define $n=|V(G)|, m=|E(G)|$, respectively. The matrix $T=(T_{u,v}) \, (u, v \in V)$ is an $n \times n$ transition matrix of isotropic RWs, that is,

\[ T_{u, v}= \begin{cases}
				1/\mathrm{deg}(u) & \text{if $u \sim v$}, \\
				0 & \text{otherwise}. \\
			 \end{cases}
				\]

For $u$, $v \in V$, $e=(u, v)$ is an arc from $u$ to $v$, and $e^{-1}=(v, u)$. We denote $o(e)=u$, $t(e)=v$. In addition, $D(G)=\{ (u, v), (v, u) | u, v \in V \}$ is a set of symmetric arcs of $G$. We consider $\dC^ {|D(G)|} =\dC^{2m}$ as a state space. For a square matrix $A$, if we note

\begin{equation*}
\mathrm{Spec}(A)= \left (
	\begin{array} {cccc}
	\lambda _{1} & \lambda _{2} & \cdots & \lambda _{r} \\
	m_{1} & m_{2} & \cdots & m_{r} \\
	\end{array}
	\right ),
\end{equation*}
it implies that the multiplicity of the eigenvalue $\lambda _{i}$ of $A$ is $m_{i}$ for $1\leq 
i\leq r$. We denote $\lambda \in \mathrm{Spec}(A)$, if $\lambda$ is an eigenvalue of $A$.

Throughout this paper, $P_{l}$, $K_{l}$ denotes the path, complete graph with $l$ vertices, respectively. The graph $C_{l}$ implies a cycle graph with $l$ vertices and we call it an even cycle if $l$ is even, otherwise an odd cycle. In this paper, we say that the graph is unicycle if it has only one cycle and the graph removed the cycle becomes a non-empty forest. So we do not call cycle graphs unicycle although they are unicycle generally. In addition we call it an even unicycle if the length of cycle is even, otherwise an odd unicycle. 

\begin{figure}[htbp]
\begin{tabular}{ccc}
		\begin{minipage}[h]{0.25\linewidth}
		\centering
		\scalebox{1.0}{
		\includegraphics[scale=1.0]{uni-example.1}
		}
		\caption{}
		\end{minipage}
		\begin{minipage}[h]{0.25\linewidth}
		\centering
		\scalebox{1.0}{
		\includegraphics[scale=0.9]{uni-example.2}
		}
		\caption{}
		\end{minipage}		
		\begin{minipage}[htbp]{0.25\linewidth}
		\centering
		\vspace{20pt}
		\includegraphics[scale=1.0]{uni-example.4}
		\vspace{5pt}
		\caption{}
		\end{minipage}
\end{tabular}
\end{figure}
We call the graphs like Figure 1, Figure 2 unicycle but do not call the graphs like Figure 3 so in this paper.
Let $K_{r, s}$ be a complete bipartite graph with two partitions with $r$ and $s$ vertices. The tree graph is defined as the graph without cycles. The girth of $G$ is denoted by $g(G)$, which means the length of the minimal cycle in $G$.

Furthermore, we assign vectors ${\vec{x}}_{e}$ to every $e \in D(G)$, where ${\vec{x}}_{e}$ s are the standard basis of the Hilbert space $\dC^{2m}$ , that is,  $({\vec{x}}_{e})_{f} = \delta _{e,f}$ for every $e, f \in D(G)$. We construct the quantum state at time $t$, $\varphi _{t} \in \dC^{2m}$ as
\[ \varphi _{t} = \Sigma_{e \in D(G)} \alpha ^{t}_{e} {\vec{x}}_{e}, \]
where $\alpha ^{t}_{e} \in \dC$, and $\Sigma _{e \in D(G)} {\mid  \alpha ^{t} _{e} \mid }^{2} = 1$. Then the finding probability of the walker on an arc $e$ at time $t$ is ${|\alpha ^{t} _{e}|} ^{2}$. Giving a $2m \times 2m$ unitary matrix $U$, we determine $\varphi _{t+1}$ as 
\[ \varphi _{t+1} = U \varphi _{t}. \] 
Hence, we can denote $\varphi _{t}$ as
\[ \varphi _{t} = U^{t} \varphi _{0} \]
using the initial state $\varphi _{0}$.

Next, we introduce Grover walk. The evolution operator of the Grover walk is the following $2m \times 2m$ unitary matrix $U=U(G)=(U _{e, f}) \, (e, f \in D(G))$:
\[ U_{e, f}=
\begin{cases}
	2/\mathrm{deg}(t(f)) & \text{if $t(f)=o(e)$ and  $f \not = e^{-1}$}, \\
	2/\mathrm{deg}(t(f)) - 1 & \text{if $f = e^{-1}$}, \\
	0 & \text{otherwise}.
\end{cases}
\]
 
The quantum waves on the arc $f$ transmits to the arc $e$ with $t(f)=o(e)$ and $f \not= e$ with a rate of $2/\mathrm{deg}(t(f))$, and reflects to the arc $f^{-1}$ with a rate of $2/\mathrm{deg}(t(f))-1$. This $U$ is called the Grover transfer matrix. We shall give an example of the Grover walk and consider the periodicity.
\newpage
We will provide an example of graph which induces a periodic Grover walk.\\
Example: 
	$G=K_{1,3}$. 
			
	\begin{figure}[hbtp]
	\centering
	\includegraphics[scale=1.5]{k13.1}
	\caption{$K_{1,3}$}
	\end{figure}
	
	\begin{equation*}
	U=U(K_{1, 3})= \left (
	\begin{array} {cccccc}
	0 & 0 & 0 & -1/3 & 2/3 & 2/3\\
	0 & 0 & 0 & 2/3 & -1/3 & 2/3\\
	0 & 0 & 0 & 2/3 & 2/3 & -1/3\\
	1 & 0 & 0 & 0 & 0 & 0\\
    0 & 1 & 0 & 0 & 0 & 0\\
    0 & 0 & 1 & 0 & 0 & 0\\
	\end{array}
	\right ).
\end{equation*}
In fact, this $U$ satisfies $U^{4}=I_{6}$, that is, 
$\varphi _{4}=U^{4} \varphi _{0}=\varphi _{0}$ 
for an arbitrary initial state $\varphi _{0}$. We say $G=K_{1, 3}$ is the graph to induce a $4$-periodic Grover walk. 
The other examples are given in Figure 5, 6, 7.
\begin{figure}[h]
\begin{tabular}{ccc}
		\begin{minipage}[h]{0.35\linewidth}
		\centering
		\scalebox{1.5}{
		\includegraphics[scale=1.0]{gra-example.1}
		}
		\caption{}
		\end{minipage}
		\begin{minipage}[h]{0.20\linewidth}
		\vspace{20pt}
		\centering
		\scalebox{1.5}{
		\includegraphics[scale=1.4]{gra-example.4}
		}
		\caption{}
		\end{minipage}		
		\begin{minipage}[h]{0.25\linewidth}
		\vspace{20pt}
	    \centering
		\scalebox{1.5}{
		\includegraphics[scale=1.3]{gra-example.3}
		}
		\caption{}
		\end{minipage}
\end{tabular}
\end{figure}
\newline These graphs induce $5, 6, 4$-periodic Grover walks, respectively. 
\subsection{Main Results}

For a positive integer $k$, inducing a $k$-periodic Grover walk implies that $U^{k}=I_{2m}$, while $U^{j} \not= I_{2m}$ for every $j$ with $j < k$ for the Grover transfer matrix $U$. These conditions can be regarded as the following spectral problem.
\begin{prop}
A graph $G$ induces a $k$-periodic Grover walk if and only if $\lambda ^{k}_{U} =1$ for every $\lambda _{U} \in \mathrm{Spec}(U)$, and there exists $\lambda_{U} \in \mathrm{Spec}(U)$ such that $\lambda^{j}_{U} \not= 1$ for every $j$ with $j < k$.
\end{prop}
So what we have to do is checking whether all the eigenvalues of $U$ satisfy the condition of the above Proposition. In order to consider it, we need the following Theorem.
\begin{thm}
	$\mathrm{(}$Emms,  Hancock,  Severini and Wilson $[3], [4]$ $\mathrm{)}$\\ 
    For the Grover transfer matrix $U$ and the transition matrix $T$ on a graph $G$, it holds 
	\[ det(\lambda _{U} I_{2m}-U) = (\lambda ^{2}_{U}-1)^{m-n} ((\lambda ^{2}_{U}+1)I_{n}-2\lambda _{U} T) \] for every $\lambda _{U} \in \dC$.
\end{thm}
So $U$ has $2n$ eigenvalues of the following form
\[ \lambda _{U} =\lambda _{T}\pm i \sqrt{1-\lambda _{T}^2}, \]
where $\lambda _{T} \in \mathrm{Spec}(T)$. The remaining $2(m-n)$ eigenvalues are $-1, 1$, which have the same multiplicities. Therefore if these two kind of eigenvalues satisfy the condition of Proposition 1.1, $G$ induces a $k$-periodic Grover walk. We characterize such graphs for $k=2, 3, 4, 5$ and obtain a necessary condition for an odd $k$. 

\begin{itemize}
	\item Theorem $2.1.$ The graph $P_{2}$ is the only graph to induce a $2$-periodic Grover walk. 
	\item Theorem $3.1.$ If $G$ induces an odd-periodic Grover walk, then $G$ is an odd cycle or an odd unicycle graph.
	\item Theorem $3.2.$ The graphs $C_{3}$, $C_{5}$ are the only graphs to induce $3$, $5$-periodic Grover walks, respectively.
	\item Theorem $4.1.$ The graphs $K_{r, s}$ are the only graphs to induce a $4$-periodic Grover walk for every $r, s \in \dN$.
\end{itemize}

This paper is organized as follows: In section 2, we mention the $2$-periodic case and prove Theorem 2.1. In section 3, we first give a necessary condition for graphs to induce an odd-periodic Grover walk and prove Theorem 3.2 with several Lemmas. In section 4, we prove Theorem $4.1$ by using a property of bipartite graphs. At the end of this paper, we summary our results and make some discussions in section 5.

\section{2-periodic Case}

Here we explain the graph to induce a $2$-periodic Grover walk, and prove Theorem 2.1 with some Lemmas. 

\subsection {Main Result}
\begin {thm}
The graph $P_{2}$ is the only graph to induce a $2$-periodic Grover walk.
\end {thm}

\subsection {Proof of Theorem 2.1}
\begin {lem}
For any $\lambda _{T} \in \mathrm{Spec}(T)$, it holds that $|\lambda _{T}| \leq 1$ and $1 \in \mathrm{Spec}(T)$.
\end {lem}

\begin {lem} $(Perron$-$Frobenius)$
If $A$ is a non-negative matrix, that is, all entries are non-negative, then the eigenvector of the maximal eigenvalue of $A$ is a non-negative vector and its multiplicity is $1$.
\end {lem} Proof of Theorem 2.1: Obviously $P_{2}$ induces a $2$-periodic Grover walk. Hence, we will prove that if $G$ induces a $2$-periodic Grover walk, then $G$ is $P_{2}$. By Proposition 1.1, for any $\lambda_{U} \in \mathrm{Spec}(U)$, it should hold that $\lambda^{2} _{U} =1$. According to Theorem 1.2, $U$ has eigenvalues of the form $\lambda _{T}\pm i \sqrt{1-\lambda _{T}^2}$, and the remaining $2(m-n)$ eigenvalues are $\pm 1$. The latter values satisfy the condition of Proposition 1.1, then the former values should satisfy it, which implies 
$\lambda _{T}\pm i \sqrt{1-\lambda _{T}^2} = \pm 1$, that is, 
$\lambda _{T} = \pm 1$. Since $T$ is a non-negative matrix and $1$ is the maximal eigenvalue of $T$, then its multiplicity is $1$ by Lemmas 2.2, 2.3. Moreover the multiplicity of $-1$ is also $1$ because of $\mathrm{Tr}(T)=0$. So we can obtain
\begin{equation}
\mathrm{Spec}(T)= \left (
	\begin{array} {cc}
	-1 & 1 \\
	1 & 1 \\
	\end{array}
	\right ).
\end{equation}
Considering the connectivity of $G$ and the summation of their multiplicities, we can obtain that $P_{2}$ is the only graph which leads (1) as a spectrum of its transition matrix. Indeed,
\begin{equation*}
U(P_{2})= \left (
	\begin{array} {cc}
	0 & 1 \\
	1 & 0 \\
	\end{array}
	\right ).
\end{equation*}
Obviously it induces a $2$-periodic Grover walk. Then such a graph is only $P_{2}$. $\Box$

\section {3, 5-periodic Case}

In this section we show that the graphs inducing an odd-periodic Grover walk should satisfy some conditions. Then we prove Theorem 3.1 and Theorem 3.2.

\subsection {Main Result}

\begin{thm}
If $G$ induces an odd-periodic Grover walk, then $G$ is an odd cycle or an odd unicycle graph.
\end{thm}

\begin{thm}
The graphs $C_{3}$, $C_{5}$ are the only graphs to induce $3, 5$-periodic Grover walks, respectively.
\end{thm}

\subsection{Proof of Theorem 3.1}

\begin{lem}
A graph $G$ is a bipartite graph if and only if it holds that
$(\lambda_{T})_{min}=-(\lambda_{T})_{max}$
for the eigenvalues of its transition matrix $T$.
\end{lem}
Furthermore using Lemma 2.2, Lemma 2.3, we can gain the following corollary:
\begin{cor}
A graph $G$ is a bipartite graph if and only if $-1 \in \mathrm{Spec}(T)$.
\end{cor}

Proof of Theorem 3.1: Let $k$ be an odd integer. We assume that $m-n > 0$. For all of the eigenvalues of $U$, $\lambda_{U}$ should satisfy $\lambda^{k}_{U}=1$. If $m-n > 0$, then at least one $-1$ is an eigenvalue of $U$ by Theorem 1.2. Since $k$ is odd, $-1$ does not satisfy the condition. Thus it should hold that $m-n \leq 0$. It follows that $m=n-1, m=n$ from the  connectivity of $G$. Then such graphs must be trees, which satisfy $m=n-1$, or cycles, unicycles, which satisfy $m=n$. From Corollary 3.4 and Theorem 1.2, trees, even cycles, and even unicycles are improper graphs since they are bipartite. Hence, an odd cycle and an odd unicycles can induce a $k$-periodic Grover walk for an odd $k$. $\Box$

\subsection{Proof of Theorem 3.2}

First, we introduce an easy Lemma and obtain a restriction of unicycle graphs which induce an odd-periodic Grover walk. We prove Theorem 3.2 with them. 

\begin{lem}
The graphs $C_{k}$ induce a $k$-periodic Grover walk.
\end{lem}

Proof: Let $A$ be an adjacent matrix of $C_{k}$, and $\lambda_{A}$ be an eigenvalue of $A$. For every $j$ with $0 \leq j \leq k$,
\[ \lambda _{A} = 2 \cos{ \frac{2 \pi }{k} j}. \]
Since $C_{k}$ is $2$-regular, it holds that
\[ T=\frac{1}{2} A. \]
Then $\lambda_{T}=\cos{(2\pi j/k)}$ for $0 \leq j \leq k$. Hence, the eigenvalues of $U$ are 
\[ \lambda _{T}\pm i \sqrt{1-\lambda _{T}^2} =e^{\pm \frac{2\pi i}{k} j}. \]
Thus $C_{k}$ induces a $k$-periodic Grover walk. $\Box$

Let $G$ be an odd unicycle graph.
\begin{prop}
If $G$ induce a $k$-periodic Grover walk, then it should hold that $g(G)\leq k-4$.
\end{prop}
Proof: Let $k$ be an odd integer and $g(G)=t$. We assume that an odd unicycle $G$ induce a $k$-periodic Grover walk. Since $G$ is a unicycle, $G$ contains the graph on Figure 8 as its subgraph.
\begin{figure}
\centering
\includegraphics[scale=1.5]{unico.4}
\caption{}
\end{figure}
Let the vertices of the cycle $C_{t}$ be $v_{1}, \cdots, v_{t}$. We define arcs $e_{i} \in D(G)$ such as
\[ e_{i}=
	\begin{cases}
	(v_{i},  v_{i+1}) & \text{if $1 \leq i \leq t-1$}, \\
	(v_{t},  v_{1}) & \text{if  $i=t$}.
	\end{cases}
	\]
Let $e$ be the arc $(v, v_{1})$. For the Grover transfer matrix $U$, $U^{k}_{e,f}$ can be written by
\[ U^{k}_{e, f}=\sum U_{e, h_{k-1}}U_{h_{k-1}, h_{k-2}} \cdots U_{h_{2}, h_{1}}U_{h_{1}, f}, \]
where $h_{1}, h_{2}, \cdots, h_{k-1} \in D(G)$ run over arcs which make the walk 
$f\rightarrow h_{1}\rightarrow h_{2}\rightarrow  \cdots \rightarrow  h_{k-1}\rightarrow e$ with the length $k$ in $G$. From its periodicity, it should hold that $U^{k}_{e,e}=1$. We consider two cases (i) $t\geq k$, (ii) $t = k-2$ and show $U^{k}_{e,e} \not= 1$ in the both of cases. 

(i) $t\geq k$. In order to obtain $U^{k}_{e,e}$, we consider the walks with length $k$ from $e$ to $e$. Since the distance between $e$ and $e$ is even and $G$ is a unicycle, we have to traverse $C_{t}$ to go from $e$ to $e$ with odd steps. This walk has the length at least $t+2$. So there are no walks from $e$ to $e$ with length $k$ since $t\geq k$. Therefore we can conclude that $U^{k}_{e,e}=0$, and it gives us a contradiction.

(ii) $t=k-2$. By the observation of (i), the walks should run through $C_{t}$ at once. Since $t=k-2$, there are only two walks such as 
\begin{eqnarray*}
e \rightarrow e^{-1} \underbrace{\rightarrow e_{1} \rightarrow e_{2} \rightarrow \cdots \rightarrow e_{t}}_{k-2\, steps} \rightarrow e, \\
e \rightarrow e^{-1} \underbrace{\rightarrow e^{-1}_{t} \rightarrow \cdots \rightarrow e^{-1}_{2} \rightarrow e^{-1}_{1}}_{k-2\, steps} \rightarrow e.
\end{eqnarray*}
Hence,
\begin{eqnarray}
U^{k}_{e, e} & = & U_{e, e_{t}} \cdots U_{e_{2}, e_{1}}U_{e_{1}, e^{-1}}U_{e^{-1}, e} \nonumber \\
			&   &+ U_{e, e^{-1}_{1}} \cdots U_{e^{-1}_{t-1}, e^{-1}_{t}}U_{e^{-1}_{t}, e^{-1}}U_{e^{-1}, e} \nonumber \\
			& = & \frac{2}{\mathrm{deg}(v_{1})} \cdots \frac{2}{\mathrm{deg}(v_{2})}\frac{2}{\mathrm{deg}(v_{1})} \left( \frac{2}{\mathrm{deg}(v)}-1 \right) \nonumber \\
			&   &+ \frac{2}{\mathrm{deg}(v_{1})} \cdots \frac{2}{\mathrm{deg}(v_{t})}\frac{2}{\mathrm{deg}(v_{1})} \left( \frac{2}{\mathrm{deg}(v)}-1 \right) \nonumber \\
		    & = & \left( \frac{2}{\mathrm{deg}(v)}-1 \right) \frac{8}{\{ \mathrm{deg}(v_{1}) \} ^{2}} \frac{2}{\mathrm{deg}(v_{2})} \cdots \frac{2}{\mathrm{deg}(v_{t})}.
\end{eqnarray}
It follows $\frac{2}{\mathrm{deg}(v)}-1>0$ and $\mathrm{deg}(v)=1$ from $U^{k}_{e,e}=1$. Therefore
\begin{equation*}
U^{k}_{e, e} = \frac{8}{\{ \mathrm{deg}(v_{1}) \} ^{2}} \frac{2}{\mathrm{deg}(v_{2})} \cdots \frac{2}{\mathrm{deg}(v_{t})}.
\end{equation*}
However it holds that $\mathrm{deg}(v_{1})\geq 3, \mathrm{deg}(v_{2}), \cdots, \mathrm{deg}(v_{t}) \geq 2$ due to the choice of the subgraph. Thus it follows that $U^{k}_{e,e}\not=1$. $\Box$

Proof of Theorem 3.2: From the previous arguments, the graphs inducing $3$, $5$-periodic Grover walks are $C_{3}$, $C_{5}$ or the other unicycle graphs, respectively. Proposition 3.6 leads the fact that the girth $g(G)$ of such unicycle graphs should be $g(G) \leq 1$ for $k=3$, and $5$. However $g(G) \geq 3$. Thus no unicycle graphs induce $3$, and $5$-periodic Grover walks. Therefore $C_{3}$, $C_{5}$ are the only graphs to induce $3$, $5$-periodic Grover walks, respectively. $\Box$

\section {4-periodic Case}

In this section we prove Theorem 4.1 by using a property of bipartite graphs.

\subsection {Main Result}
\begin {thm}
The graphs $K_{r, s}$ are the only graphs to induce a $4$-periodic Grover walk for every $r, s \in \dN$.
\end {thm}

\subsection {Proof of Theorem 4.1}

Similar to the previous section, we introduce some Lemmas to prove Theorem 4.1.

\begin{lem}
A graph $G$ induces a $4$-periodic Grover walk if and only if the spectrum of its transition matrix $T$ is of the form
\begin{equation}
\mathrm{Spec}(T)= \left (
	\begin{array} {ccc}
	-1 & 0 & 1 \\
	1 & n-2 & 1\\
	\end{array}
	\right ).
\end{equation}
\end{lem}

Proof: First, we show its necessity. In other words, we show that if $G$ induces $4$-periodic Grover walk, then spectrum of $T$ is of the form (3). Similar to Proof of Theorem 2.1, for any $\lambda_{U} \in \mathrm{Spec}(U)$, it should hold that $\lambda^{4} _{U} =1$, and it follows that
\[ \lambda _{T}\pm i \sqrt{1-\lambda _{T}^2}= \pm 1, \pm i. \] Thus, $\lambda_{T}= \pm 1, 0$. Therefore we can obtain
\begin{equation*}
\mathrm{Spec}(T)= \left (
	\begin{array} {ccc}
	-1 & 0 & 1 \\
	1 & n-2 & 1\\
	\end{array}
	\right ).
\end{equation*}
Immediately, their multiplicities are determined by Lemma 2.2 and Lemma 2.3. 

Next we show its sufficiency. If the spectrum of $T$ of $G$ is of form (3), then the spectrum of $U$ is of the form
\begin{equation*}
\mathrm{Spec}(U)= \left (
	\begin{array} {cccc}
	-1 & -i & i & 1 \\
	m-n+2 & n-2 & n-2 & m-n+2\\
	\end{array}
	\right ).
\end{equation*}
From Proposition 1.1, $G$ induces a $4$-periodic Grover walk. $\Box$\\\\
In fact this spectrum induces complete bipartite graphs immediately. 
\begin{lem}
A graph $G$ is a complete bipartite graph if and only if the spectrum of its transition matrix $T$ is of the form
\begin{equation}
\mathrm{Spec}(T)= \left (
	\begin{array} {ccc}
	-1 & 0 & 1 \\
	1 & n-2 & 1\\
	\end{array}
	\right ).
\end{equation}
\end{lem}

Proof: First, we show its necessity. If $G$ is a complete bipartite graph, then its transition matrix $T$ can be written by

\begin{equation*}
T= \left (
	\begin{array} {cc}
	O_{r, r} & \frac{1}{s} J_{r, s} \\
	\frac{1}{r} J_{s, r} & O_{s, s} \\
	\end{array}
	\right ),
\end{equation*}
where $O_{i, j}$,  $J_{i, j}$ denote the $i \times j$ matrix with all $0$ entries, and the $i \times j$ matrix with all $1$ entries, respectively. Then it follows that $\mathrm{rank}(T)=2$, and $\mathrm{dim}(\mathrm{ker}(T))=n-2$. It implies that the multiplicity of an eigenvalue $0$ of $T$ is $n-2$ since $T$ is a diagonalizable matrix. Furthermore $1$ is an eigenvalue of $T$ with multiplicity $1$, and $-1$ is so from Corollary 3.4. Therefore we can obtain (4) as the  spectrum of the transition matrix of $G$.

Next we show its sufficiency. If the spectrum of $T$ is of the form (4), then $-1$ is an eigenvalue of $T$. Thus $G$ is a bipartite graph by Corollary 3.4. Then its transition matrix $T$ can be represented by 
\begin{equation*}
T= \left (
	\begin{array} {cc}
	O_{r, r} & V_{r, s} \\
	W_{s, r} & O_{s, s} \\
	\end{array}
	\right )
\end{equation*}
with some $r \times s$ matrix $V_{r,s}$, and $s \times r$ matrix $W_{s,r}$. These matrices are  the transition matrices from a partition $R$ to another partition $S$, and from $S$ to $R$, respectively, where $V(G)=R \sqcup S$ and $r=|R|$, $s=|S|$. In addition each summations over the row of $T$ are $1$. We show that all of the entries of $V_{r,s}$, and $W_{s,r}$ are not $0$.  From the assumption of $\mathrm{dim}(\mathrm{ker}(T))=n-2$, we can see that $\mathrm{rank}(T)=2$. Therefore $V_{r,s}$ and $W_{s,r}$ can be represented by 
\begin{equation*}
V_{r,s}= \left (
	\begin{array} {c}
	c_{1} \vec{a} \\
	\vdots \\
	c_{r} \vec{a} \\
	\end{array}
	\right ), 
\end{equation*}
\begin{equation*}
W_{s,r}= \left (
	\begin{array} {c}
	d_{1} \vec{b} \\
	\vdots \\
	d_{s} \vec{b} \\
	\end{array}
	\right ), 
\end{equation*}
where $\vec{a}$ is a $1 \times r$ rational vector and $\vec{b}$ is a $1 \times s$ rational vector, and $c_{i}, d_{j}$ are rational numbers for  $1\leq i\leq r$, $1\leq j\leq s$. We can obtain $c_{i}=1$, $d_{j}=1$ for $1\leq i\leq r$, $1\leq j\leq s$ since the summations over the row of $T$ are $1$. If there exists $i$ such that the $i$-th entry of $\vec{a}$ is $0$, then the $i$-th column vector of $T$ is the zero vector. It contradicts the connectivity of $G$. Thus the $V_{r,s}$ does not contain $0$ as its entries. Similarly the $W_{s,r}$ does not so. Therefore any vertices in $R$ are adjacent to the vertices in $S$, and the vertices in $S$ are so. We can conclude that $G$ is a complete bipartite graph. $\Box$

Proof of Theorem 4.1: From Lemma 4.2, $G$ induces a $4$-periodic Grover walk if and only if the spectrum of its transition matrix is of the form (4). Having such a spectrum of $T$ means that $G$ is a complete bipartite graph by Lemma 4.3. Then we can show that the complete bipartite graphs $K_{r,s}$ are the only graphs to induce a $4$-periodic Grover walk. $\Box$

\section{Summary and Discussions}

In this paper, we have given some characterization of the graphs to induce a $k$-periodic Grover walk for $k=2, 3, 4, 5$. We proved that $P_{2}, C_{3}, K_{r, s}, C_{5}$ are the only graphs to induce $2, 3, 4, 5$-periodic Grover walks, respectively. One of what we want to do is to determine such graphs for an integer $k$ with $k \geq 6$. The main method used in this paper to characterize such graphs is to analyze the spectrum of its transition matrix, that is, we found graphs whose spectrum of its transition matrix is occupied by the real part of the $k$-th root of $1$ and have an eigenvalue that is not the real part of $j$-th root of $1$ for every $j$ with $j<k$. Generally speaking, it is difficult to characterize graphs with any given spectrum. For the cases of $k \geq 6$, we might take another method to solve it. 

Next, we will provide some examples to induce a $k$-periodic Grover walk for $k \geq 6$ and a special operator between graphs. 

\begin{lem}
The graphs $P_{k}$ induce a $2(k-1)$-periodic Grover walk.
\end{lem}
Using this Lemma and Lemma 3.5, we can conclude that $P_{4}$ and $C_{6}$ induce $6$-periodic Grover walks. In addition both of the graphs on Figure 9, 10 also induce $6$-periodic Grover walks. These graphs are made by identifying two endpoints of some $P_{4}$ s. They include $P_{4}$ and $C_{6}$. Furthermore both of the graphs on Figure 11, 12 induce $8$-periodic Grover walks.

\begin{figure}[h]
\centering
\begin{tabular}{cc}
		\begin{minipage}[h]{0.25\linewidth}
		\centering
		\scalebox{1.0}{
		\includegraphics[scale=1.0]{6peri.1}
		}
		\caption{}
		\end{minipage}
		\begin{minipage}[h]{0.25\linewidth}
		\centering
		\scalebox{1.0}{
		\includegraphics[scale=1.0]{6peri2.2}
		}
		\caption{}
		\end{minipage}
\end{tabular}
\end{figure}

\begin{figure}[h]
\centering
\begin{tabular}{cc}
		\begin{minipage}[h]{0.25\linewidth}
		\centering
		\scalebox{1.0}{
		\includegraphics[scale=1.0]{8peri.1}
		}
		\caption{}
		\end{minipage}
		\begin{minipage}[h]{0.25\linewidth}
		\centering
		\scalebox{1.0}{
		\includegraphics[scale=1.0]{8peri2.2}
		}
		\caption{}
		\end{minipage}
\end{tabular}
\end{figure}
These graphs include $P_{5}$, $C_{8}$. Then the graphs to induce an even-periodic Grover walk contain these graphs, and we must find any other graphs. 

For an odd $k$, it is thought that $C_{k}$ is the only graph which induce a $k$-periodic Grover walk. So we have to eliminate the possibility of the unicycle graphs. 

Moreover we get the following Proposition.
\begin{prop}
If $G$ induces a $k$-periodic Grover walk then its subdivision graph $S(G)$ induces a $2k$-periodic Grover walk. 
\end{prop}

\begin{figure}[h]
\begin{tabular}{cc}
		\begin{minipage}[h]{0.25\linewidth}
		\centering
		\scalebox{1.0}{
		\includegraphics[scale=1.5]{sp-gra.3}
		}
		\caption{$G$}
		\end{minipage}
		\begin{minipage}[h]{0.25\linewidth}
		\scalebox{1.0}{
		\includegraphics[scale=1.5]{sp-gra.4}
		}
		\caption{$S(G)$}
		\end{minipage}
\end{tabular}
\end{figure}

$G, S(G)$ induce $4, 8$-periodic Grover walks, respectively. We can regard the subdivision as an operator which conserves the periodicity of the Grover walk between graphs. To find such operators between graphs is also our interest. The Grover walk on the graphs are determined by the Grover transfer matrix. We have to investigate the periodicity of another QWs on the graphs determined by a unitary matrix except the Grover transfer matrix.

\end {document}